\documentstyle[12pt]{article}
\newcommand{\prd}[3]{{\it Phys. Rev.} {\bf D#1}, #2 (19#3)}
\newcommand{\pl}[3]{{\it Phys. Lett.} {\bf #1B}, #2  (19#3)}
\newcommand{\np}[3]{{\it Nucl. Phys.} {\bf B#1}, #2  (19#3)}
\newcommand{\prl}[3]{{\it Phys. Rev. Lett.} {\bf #1}, #2  (19#3)}

\newcommand{\ra}{\rightarrow}

\newcommand{\matel}[3]{\left<#1\right|#2\left|#3\right>}
\newcommand{\cp}{{\cal CP}}

\def\uu{\overline{u}}
\def\dd{\overline{d}}

\def\rrr{{B_0^2\over m_K^2}}
\def\ff{\overline{\Psi}}
\def\ll{\lambda^a}
\def\gm{\gamma_\mu}
\def\kob{\overline{K}^0}
\relax
\begin{document}
\begin{titlepage}
\def\ba{\begin{array}}
\def\ea{\end{array}}
\def\thefootnote{\fnsymbol{footnote}}
\vfill
\hskip 4in ISU-HET-95-3

\hskip 4in OITS-579

\hskip 4in June 1995
\vspace{1 in}
\begin{center}
{\large \bf $\cp$ Violation in $\Lambda \ra p \pi^-$\\
 Beyond the Standard Model}\\

\vspace{1 in}
{\bf  Xiao-Gang~He$^{(a)}$ and G.~Valencia$^{(b)}$}\\
{\it $^{(a)}$~Institute of Theoretical Science,
             University of Oregon, Eugene, OR 97403}\\
{\it $^{(b)}$~Department of Physics,
               Iowa State University,
               Ames, IA 50011}\\
\vspace{1 in}
     %	{\large \bf ABSTRACT}
\end{center}

\begin{abstract}

The $\cp$ violating asymmetry $A(\Lambda^0_-)$ has been estimated to
occur at the level of a few times $10^{-5}$
within the minimal standard model. The
experiment E871 expects to reach a sensitivity of $10^{-4}$ to the
asymmetry $A(\Lambda^0_-)+A(\Xi^-_-)$. In this paper we study some of the
implications of such a measurement for $\cp$ violation beyond the minimal
standard model. We find that it is possible to have $A(\Lambda^0_-)$ at the
few times $10^{-4}$ level while satisfying the constraints imposed by
the measurements of $\cp$ violation in kaon decays.

\end{abstract}

\end{titlepage}

\section{Introduction}

The origin of $\cp$ violation remains one of the outstanding problems in
particle physics. In the attempt to understand this problem many experimental
and theoretical efforts have been launched \cite{review}. One of the systems
where it is possible to search for $\cp$ violation is the non-leptonic
decay of hyperons. Although this has been known for many years \cite{marshak},
it is only recently that it has become conceivable to carry out an experimental
program to look for $\cp$ violating signals in the decays of $\Xi$ and
$\Lambda$ hyperons \cite{cernpro,proposal}.

Of particular interest is the upcoming experiment E871 that expects to
reach a sensitivity of $10^{-4}$ for the sum of asymmetries $A(\Lambda^0_-)
+ A(\Xi^-_-)$ \cite{proposal}.
Unfortunately, the calculation of these asymmetries is
plagued by theoretical uncertainties in the estimate of the hadronic matrix
elements involved. Nevertheless, a conservative study of these asymmetries
within the minimal standard model indicated that $A(\Lambda^0_-)$ is likely to
occur at the level of a few times $10^{-5}$. In view of this, the potential
results of E871 are very exciting.

One of the questions we would like to answer is whether the phase in the CKM
matrix of the three generation minimal standard model is the sole source of
$\cp$ violation. The experimental information that we have so far is:
\begin{itemize}
\item A
non-zero value of the parameter $\epsilon$ in kaon decays \cite{pdb}:
\begin{equation}
|\epsilon| = 2.26 \times 10^{-3}
\end{equation}

\item A measurement
of the parameter $\epsilon^\prime$ \cite{epove}:
\begin{equation}
{\epsilon^\prime \over \epsilon} = \cases{ (2.3 \pm 0.65)\times 10^{-3} &
NA31 \cr
(0.74\pm0.52\pm0.29)\times 10^{-3} & E731 \cr}
\label{eppexp}
\end{equation}
\end{itemize}
The first result indicates that there is $\cp$
violation in nature, but it does not
pinpoint its origin. The best one can say is that it is possible for the
minimal standard model to accommodate this number.
If the second number turns out to
be non-zero it would establish the existence of
direct $|\Delta S|=1$ $\cp$ violation, ruling out some superweak models.
The current experimental numbers are
consistent with the minimal standard model, although the theoretical
calculations are also plagued with uncertainty from the
evaluation of hadronic matrix elements.

The present situation is, therefore, that there is no need for $\cp$ violation
beyond the phase in the three generation CKM matrix\footnote{Except perhaps
in the origin of the baryon asymmetry of the universe. We will not discuss that
issue in this paper.}, but that other sources of $\cp$ violation have not
been ruled out.

The question we want to address in this paper is whether it
is possible for E871 to find a non-zero asymmetry given its expected
sensitivity and the current values of $\epsilon$ and
$\epsilon^\prime/\epsilon$. To this end, and in keeping with the
results of all the precise experiments conducted to date,  we will assume
that the minimal three-generation standard model is a very good low energy
approximation to the electroweak interactions. We will, therefore, discuss
any possible new physics in terms of an effective Lagrangian consistent
with the symmetries of the standard model and will only look only at
operators of dimension six.

Our paper is organized as follows. In Section 2 we review the notation
for $\cp$ violating observables in hyperon decays as well as the
standard model estimate of $A(\Lambda^0_-)$. In section~3 we compute
the contributions of $\cp$ violating four quark operators to
$A(\Lambda^0_-)$ and the constraints that result from the measurements
of $\cp$ violation in $K\ra \pi\pi$. In section~4 we repeat this
analysis for the two-quark operators of dimension six (so called penguin
operators). Finally, we present our conclusions.

\section{$\cp$ Violation in $\Lambda^0 \ra p \pi^-$}

In this section we review the basic features of $\cp$ violation in
the reaction $\Lambda^0 \ra p \pi^-$, denoted by $(\Lambda^0_-)$.
In the $\Lambda^0$ rest frame, $\vec{\omega}_{i,f}$ will denote unit vectors in
the directions of the $\Lambda$ and $p$ polarizations, and
$\vec{q}$ will denote the proton momentum.
The isospin of the final state is
$I= 1/2 {\rm ~or~} 3/2$, and each of these two states can be reached
via a $\Delta I = 1/2 {\rm ~or~} 3/2$ weak transition respectively.
There are also two possibilities for the parity of the final state.  They are
the $s$-wave, $l=0$, parity odd state (thus reached via a parity
violating amplitude); and the $p$-wave, $l=1$, parity even state reached
via a parity conserving amplitude.

A model independent analysis of the decay can be done by writing the
most general matrix element consistent with Lorentz
invariance: \cite{marshak}
\begin{equation}
{\cal M} = G_F m_\pi^2 \overline{u}_P(A-B\gamma_5)u_\Lambda .
\label{gmatel}
\end{equation}
It is customary to introduce the quantities:
\begin{eqnarray}
s&=&A\nonumber \\
p&=&\biggl({|\vec{q}|\over E_P + M_P}\biggr) B
\end{eqnarray}
to write the total decay:
\begin{equation}
\Gamma = {|\vec{q}|(E_P+M_P) \over 4 \pi M_\Lambda}G_F^2 m_\pi^4
\left(|s|^2+|p|^2\right).
\label{drate}
\end{equation}
The angular distribution is proportional to:
\begin{equation}
{d\Gamma \over d\Omega} \sim 1+
\gamma\vec{\omega}_i\cdot\vec{\omega}_f + (1-\gamma)\hat{q}\cdot\vec{\omega}_i
\hat{q}\cdot\vec{\omega}_f
+ \alpha \hat{q}\cdot(\vec{\omega}_i+\vec{\omega}_f)
+ \beta \hat{q}\cdot(\vec{\omega}_f\times\vec{\omega}_i),
\label{msqsim}
\end{equation}
where we have used the standard notation \cite{marshak}:
\begin{equation}
\alpha  \equiv {2 {\rm Re} s^*p \over |s|^2 + |p|^2},  \;\;
\beta  \equiv {2 {\rm Im} s^*p \over |s|^2 + |p|^2}, \;\;
\gamma  \equiv {|s|^2 - |p|^2 \over |s|^2 + |p|^2} .
\label{albega}
\end{equation}
If the proton polarization is not observed, $\alpha$ is the parameter
that governs the angular distribution:
\begin{equation}
{d\Gamma \over d\Omega}={\Gamma \over 4 \pi}
\left(1+\alpha \hat{q}\cdot \vec{\omega}_i\right).
\label{alonly}
\end{equation}
Similarly, if the initial $\Lambda$ is unpolarized,
$\alpha$ determines the polarization of the proton:
\begin{equation}
\vec{\cal P}_p = \alpha_\Lambda \hat{q}
\label{otheral}
\end{equation}
E871 will not measure the correlations governed by the
parameter $\beta$ so we will not deal with it in this paper.

The $\cp$-odd observable $A(\Lambda^0_-)$ is constructed by
comparing the parameter $\alpha$ in the reaction
$\Lambda^0 \ra p \pi^-$ with the corresponding parameter
$\overline{\alpha}$ in the reaction
$\overline{\Lambda}^0 \ra \overline{p} \pi^+$.
One can show that $\cp$ symmetry predicts that:
\begin{equation}
\overline{\alpha} = - \alpha
\label{cppred}
\end{equation}
so that a $\cp$ odd observable is:
\cite{donpa}
\begin{equation}
A  \equiv {\alpha \Gamma + \overline{\alpha}\overline{\Gamma} \over
\alpha \Gamma - \overline{\alpha}\overline{\Gamma}}
\approx {\alpha + \overline{\alpha} \over \alpha - \overline{\alpha}}
\label{cpobs}
\end{equation}
Other possible $\cp$ odd observables have been discussed in the literature:
a rate asymmetry that is significantly smaller than $A$ \cite{donpa}; and
an asymmetry based on the parameter $\beta$ that won't be accessible to E871.
For these reasons we concern ourselves with the observable $A(\Lambda^0_-)$.
\footnote{In fact E871 will be sensitive to the
sum $A(\Lambda^0_-) + A(\Xi^-_-)$. An analysis of
$A(\Xi^-_-)$ parallels the one we will
carry out, but doesn't really affect our conclusions given the inherent
uncertainties in the computation of matrix elements. It has also been argued
that $A(\Xi^-_-)$ is probably smaller than $A(\Lambda^0_-)$ due to much smaller
strong rescattering phases \cite{lusawi}.}

It is convenient to decompose the amplitudes according to isospin, and
to introduce the following notation for the phases:
\begin{eqnarray}
s(\Lambda^0_-)= - \sqrt{2/3}s_{1}e^{i(\delta^s_1+\phi^s_1)}+\sqrt{1/3}
s_{3}e^{i(\delta^s_3+\phi^s_3)}\nonumber \\
p(\Lambda^0_-)= - \sqrt{2/3}p_{1}e^{i(\delta^p_1+\phi^p_1)}+\sqrt{1/3}
p_{3}e^{i(\delta^p_3+\phi^p_3)}
\label{overnot}
\end{eqnarray}
where $\delta^I_J$ is the strong rescattering phase for the pion nucleon
system and $\phi^I_J$ is the $\cp$ violating  phase.

In terms of these quantities one finds: \cite{donpa}
\begin{eqnarray}
A(\Lambda^0_-)&=& -\tan\left(\delta^p_1-\delta^s_1\right)
\sin\left(\phi^p_1-\phi^s_1\right)\biggl[ 1 \nonumber \\
&+&{1\over \sqrt{2}}{s_3 \over  s_1}\biggl( {\cos(\delta^p_1-\delta^s_3)
\over \cos (\delta^p_1-\delta^s_1)}-{\sin(\delta^p_1-\delta^s_3)
\over \sin(\delta^p_1-\delta^s_1)}{\sin(\phi^p_1-\phi^s_3)
\over \sin(\phi^p_1-\phi^s_1)}\biggr) \nonumber \\
&+&{1\over \sqrt{2}}{p_3 \over p_1}\biggl( {\cos(\delta^p_3-\delta^s_1)
\over \cos (\delta^p_1-\delta^s_1)}-{\sin(\delta^p_3-\delta^s_1)
\over \sin(\delta^p_1-\delta^s_1)}{\sin(\phi^p_3-\phi^s_1)
\over \sin(\phi^p_1-\phi^s_1)}\biggr) \biggr]
\label{approxas}
\end{eqnarray}

Experimentally we know the values of:
\begin{itemize}
\item the strong rescattering phases \cite{roper}:
\begin{equation}
\delta_1 \approx 6.0^\circ,\;\;\delta_3 \approx -3.8^\circ,\;\;
\delta_{11}\approx -1.1^\circ,\;\;\delta_{31}\approx-0.7^\circ
\label{stphex}
\end{equation}
with all the errors on the order of $1^\circ$.

\item the $\Delta I=3/2$ amplitudes are much smaller than the
$\Delta I=1/2$ amplitudes \cite{pdbee}:
\begin{equation}
s_{3}/s_{1} = 0.027 \pm 0.008, \;\; p_{3}/p_{1}=0.03\pm 0.037
\label{expamp}
\end{equation}

\item the $s$ and $p$ amplitudes (assuming they are dominated by the
$\cp$ conserving, $\Delta I =1/2$, transitions):
\begin{eqnarray}
s \approx -\sqrt{2 \over 3} s_1 &=& 1.47 \pm 0.01 \nonumber \\
p \approx -\sqrt{2 \over 3} p_1 &=& \biggl({|\vec{q}|\over E_P + M_P}\biggr)
(9.98 \pm 0.24)
\end{eqnarray}

\end{itemize}
Substituting the experimental numbers for the amplitudes and strong
rescattering phases one gets:
\begin{equation}
A(\Lambda^0_-)\approx 0.13 \sin(\phi^p_1-\phi^s_1)
+0.001 \sin(\phi^p_1-\phi^s_3) -0.0024 \sin(\phi^p_3-\phi^s_1)
\label{alambda}
\end{equation}

\subsection{Standard model calculation}

In the case of the minimal standard model, the $\cp$ violating phase resides
in the CKM matrix. For low energy transitions, this phase shows up as the
imaginary part of the Wilson coefficients in the effective weak
Hamiltonian. In the notation of Buras \cite{buras},
\begin{equation}
H_W^{SM} = {G_F \over \sqrt{2}}V^*_{ud}V_{us}\sum_i c_i(\mu)Q_i(\mu) +
{\rm ~h.~c.}
\label{effweak}
\end{equation}
$Q_i(\mu)$ are four quark operators, and $c_i(\mu)$ are the Wilson
coefficients that are usually written as:
\begin{eqnarray}
c_i(\mu) &=&z_i(\mu)+\tau y_i(\mu) \nonumber \\
\tau &=& - {V^*_{td}V_{ts} \over V^*_{ud}V_{us}}
\end{eqnarray}
with the $\cp$ violating phase being the phase of $\tau$.
Numerical values for these coefficients can be found, for
example, in Buchalla {\it et. al.} \cite{buras}.

The calculation of the weak phases would proceed
by evaluating the hadronic matrix
elements of the four-quark operators in Eq.~\ref{effweak} to obtain real
and imaginary parts for the amplitudes, schematically:
\begin{equation}
\matel{p \pi}{H_w^{eff}}{\Lambda^0}|^I_\ell = {\rm Re}M^I_\ell + i {\rm Im}
M^I_\ell ,
\label{schematic}
\end{equation}
and to the extent that the $\cp$ violating phases are small, they can be
approximated by
\begin{equation}
\phi^I_\ell \approx { {\rm Im}M^I_\ell \over {\rm Re}M^I_\ell}.
\label{smallph}
\end{equation}
At present, however, we do not know how to compute the matrix elements so
we cannot actually implement this calculation.

For a simple estimate, we can take the real part of the matrix
elements from experiment (assuming that the measured amplitudes are real,
that is, that $\cp$ violation is small), and compute the imaginary
parts in vacuum saturation. This approach provides a conservative
estimate for the weak phases because the model calculation of the
real part of the amplitudes is smaller than the experimental value.
Nevertheless, the numbers should be viewed with great caution.

The approximate weak phases estimated in vacuum saturation are: \cite{steger}
\begin{eqnarray}
\phi^1_s &\approx& -3 y_6 {\rm Im}\tau \nonumber \\
\phi^1_p &\approx& -0.3 y_6 {\rm Im}\tau \nonumber \\
\phi^3_s &\approx& \left[3.6(y_1+y_2)+2.7(y_7+3y_8)
{B_0^2 \over m_K^2}\right] {\rm Im}\tau \nonumber \\
\phi^3_p &\approx& \left[0.5(y_1+y_2)-0.4(y_7+3y_8)
{B_0^2 \over m_K^2}\right] {\rm Im}\tau
\label{approxph}
\end{eqnarray}
These provide numerical estimates using the values for the
Wilson coefficients\footnote{For
$\mu=1$~GeV, $\Lambda_{QCD}=200$~MeV}
of Buchalla {\it et. al.} \cite{buras},
$y_6 \approx -0.08$; and the value of $B_0$ given in the
appendix. For the quantity ${\rm Im}\tau$ (we use the Wolfenstein
parameterization of the CKM matrix) we take:
\begin{equation}
{\rm Im}\tau = A^2 \lambda^4 \eta \leq 0.001
\label{imtau}
\end{equation}
Putting all the numbers together, and using the upper limit
in Eq.~\ref{imtau}  yields:\footnote{See Ref.~\cite{steger} for
additional discussions of this calculation.}
\begin{equation}
A(\Lambda^0_-) \approx 3 \times 10^{-5}
\label{vsnumres}
\end{equation}

Other models of $\cp$ violation contain additional short distance operators
with $\cp$ violating phases \cite{wein,moha,senj} and predict different values
for $A(\Lambda^0_-)$ \cite{donpa,chang}.
A summary of results can be found, for example, in Ref.~\cite{hedpf}.

\section{Four-quark Operators}

We now study, in a model independent manner, the
contributions to $A(\Lambda^0_-)$ that occur
due to physics beyond the minimal standard model. In this section
we look at the effect of all the four-quark operators and in the next
section we discuss the two-quark operators.
We assume that the physics that lies beyond the minimal
standard model is characterized by a scale $\Lambda \gg M_W$ and, therefore,
that its most important low energy effects can be parameterized by the lowest
dimension operators of the most general effective Lagrangian consistent
with the symmetries of the standard model. Such a Lagrangian has been written
down by Buchm\"{u}ller and Wyler \cite{buchmuller}. In the appendix we list
all the operators that occur at dimension six with $|\Delta S| =1$.

The calculation then proceeds as in the previous section,
but with the effective Hamiltonian
\begin{equation}
H_{eff} = H_W^{SM} + {g^2 \over \Lambda^2}
\biggl(\sum_i\lambda_i {\cal O}^{new}_i + {\rm ~h.~c.}\biggr)
\label{effh}
\end{equation}
To the usual, QCD corrected, standard weak Hamiltonian of the previous section
we add all the four-fermion operators with $|\Delta S|=1$ that come from
the new physics sector. We will sidestep the issue of
the possible origin of the effective $\cp$ violating operators.
We use the notation of Ref.~\cite{buchmuller} as
detailed in the Appendix. These
operators violate $\cp$ if the coupling $\lambda_i$ has an imaginary
part. The normalization has been chosen for convenience.

\subsection{$K_L \ra \pi \pi$ and $\epsilon^\prime / \epsilon$}

The standard notation for the $K \ra \pi \pi$ amplitudes is:
\begin{eqnarray}
A(K^0 \ra \pi^+ \pi^-) &=& A_0 e^{i\delta_0}+{A_2 \over \sqrt{2}}
e^{i\delta_2} \nonumber \\
A(K^0 \ra \pi^0 \pi^0)&=& A_0 e^{i\delta_0}- \sqrt{2}A_2
e^{i\delta_2}
\end{eqnarray}
where $\delta_{0,2}$ are the strong $\pi\pi$ scattering phases in the
$I=0,2$ channel. The amplitudes $A_0$ and $A_2$ are real unless
there is $\cp$ violation.
Experimentally it is known that the $\Delta I = 3/2$ amplitude $A_2$
is much smaller than the $\Delta I = 1/2$ amplitude $A_0$:
\begin{equation}
\omega \equiv {{\rm Re}A_2 \over {\rm Re}A_0} \approx {1 \over 22}
\end{equation}
The contribution of the dominant penguin operator (${\cal O}_6$ in the notation
of Ref.~\cite{buras}) to $\epsilon^\prime/\epsilon$ is given by:
\begin{equation}
\biggl({\epsilon^\prime \over \epsilon }\biggr)_6
= -{\omega \over \sqrt{2} |\epsilon|}
{{\rm Im}(A_0)_6 \over |A_0|}=
{\omega \over 2 |\epsilon|}{G_F \over |A_0|} y_6 \lambda {\rm Im}\tau
<\pi^+\pi^-|{\cal O}_6|K^0>
\end{equation}
The hadronic uncertainty enters the calculation through the matrix
element of the four-quark operator. We will use the estimate
of Ref.~\cite{buras}
for the matrix element of ${\cal O}_6$:
\begin{equation}
<\pi^+\pi^-|{\cal O}_6|K^0>\bigg|_{I=0} = -4 \sqrt{2}f_\pi
{m_K^2 - m_\pi^2 \over
\Lambda_\chi^2}\biggl({m_K^2 \over m_s + m_d}\biggr)^2 \approx
-0.26 {\rm GeV}^3
\end{equation}
Using the values $A=0.9$, $\lambda =0.22$ and $\eta =0.5$ one finds that:
\begin{equation}
\biggl({\epsilon^\prime \over \epsilon}\biggr)_6\approx 1.5 \times 10^{-3}
\end{equation}

The usual standard model analysis of $\epsilon^\prime/\epsilon$ consists of
computing this contribution of the ``penguin'' operator,
and of normalizing all
other contributions to it in terms of a parameter $\Omega$:
\begin{equation}
{\epsilon^\prime \over \epsilon}=\biggl({\epsilon^\prime \over \epsilon}
\biggr)_6 \biggl( 1 -\Omega_{SM}-\Omega_{NEW}\biggr)
\end{equation}
$\Omega_{SM}$ is given, for example, in Ref.~\cite{buras}, and we have
introduced an analogous term $\Omega_{NEW}$ for the contributions of
the new four-quark operators. Given the experimental result
in Eq.~\ref{eppexp},
we will place bounds on the new physics by requiring, conservatively, that
$\Omega_{NEW}\leq 1$. We find:
\begin{eqnarray}
\Omega_{NEW} &=& 8\biggl({M_W \over \Lambda}\biggr)^2\sum_i
\biggl({ {\rm Im}\lambda_i
\over A^2 \lambda^5 \eta}\biggr)\biggl[
{<\pi^+\pi^-|{\cal O}_i|K^0>_{I=0}\over
y_6 <\pi^+\pi^-|{\cal O}_6|K^0>_{I=0} } \nonumber \\
&& - {\sqrt{2}\over \omega}
{<\pi^+\pi^-|{\cal O}_i|K^0>_{I=2}\over
y_6 <\pi^+\pi^-|{\cal O}_6|K^0>_{I=0} } \biggr]
\label{epprime}
\end{eqnarray}
Because there is no way at present to compute the matrix elements of
four-quark operators reliably, we will simply use vacuum saturation.
The new contributions to $\epsilon^\prime$ can thus be computed with the
aid of the matrix elements listed in Table~\ref{t: matelkpp}.
We use, as before, $A^2\lambda^5\eta \approx 2 \times 10^{-4}$ and we
explicitly separate the contributions from the different isospin
components of each operator for later convenience. We thus write:
\begin{equation}
\Omega_{NEW} = 4 \times 10^4
\biggl({M_W \over \Lambda}\biggr)^2\sum_i {\rm Im}\lambda_i
 \biggl( \omega_{0i} + \omega_{2i} \biggr)
\label{numcoe}
\end{equation}
where $\omega_{0,2i}$ refers to the $\Delta I=1/2,3/2$ component of
${\cal O}_i$. We present numerical results for $\omega_{0,2i}$ in
Table~\ref{t: omega}.
\begin{table}[tbh]
\centering
\caption[]{Numerical coefficients for Eq.~\ref{numcoe}}
\begin{tabular}{|c|c|c|} \hline
Operator &  $\omega_0$  & $\omega_2$ \\ \hline
% & & \\
${\cal O}_{qq}^{(1,1)}$ & $-0.06$ & 0 \\
${\cal O}_{qq}^{(8,1)}$ & $-0.3$ & 0 \\
${\cal O}_{qq}^{(1,3)}$ & $-0.3$ & 0 \\
${\cal O}_{qq}^{(8,3)}$ & $0.32$ & 0 \\ \hline
${\cal O}_{dd}^{(1)}$ & $0.08$ & $2.5$ \\
${\cal O}_{ud}^{(1)}$ & $-0.02$ & $-2.5$ \\
${\cal O}_{dd}^{(8)}$ & $0.1$ & $3.3$ \\
${\cal O}_{ud}^{(8)}$ & $0.2$ & $-3.3$ \\ \hline
${\cal O}_{qu}^{(1)}$ & $2.4$ & $-36.8$ \\
${\cal O}_{qu}^{(8)}$ & $0.1$ & $3.3$ \\
${\cal O}_{qd}^{(1)}$ & $1.5$ & 0 \\
${\cal O}_{qsd}^{(1)}$ & $-3.9$ & $36.8$ \\
${\cal O}_{qsd}^{(8)}$ & $-0.1$ & $-3.3$ \\ \hline
${\cal O}_{qsq}^{(1)}$ & $1.8$ & $-31.2$ \\
${\cal O}_{qsq}^{(8)}$ & $-3.5$ &$33$ \\
${\cal O}_{qqs}^{(1)}$ & $-3.5$ & $31$ \\
${\cal O}_{qqs}^{(8)}$ & $2.1$ & $-33$ \\
${\cal O}_{qq}^{(1s)}$ & $1.8$ & 0 \\
${\cal O}_{qq}^{(8s)}$ & $1.3$ & 0
\\ \hline
\end{tabular}
\label{t: omega}
\end{table}

Requiring that $\Omega_{NEW} < 1$, we can constrain the size of the
$\cp$ violating couplings Im~$\lambda_i/\Lambda^2$. By assuming
that there is no accidental cancellation between the contributions
of different operators to $\Omega_{NEW}$ we may constrain each
operator separately. The isospin decomposition is useful because
it is possible to construct combinations of operators with definite
isospin transformation properties. The constraints that apply to
operators that are purely $\Delta I =1/2$ are different from those
that apply to operators that are purely $\Delta I = 3/2$.

\subsection{$K^0-\kob$ Mixing and $\epsilon$}

In general, $\epsilon^\prime$ provides tighter constraints on new
$\cp$ violating interactions that does $\epsilon$. Nevertheless,
it is necessary to consider constraints from $\epsilon$ because
the ones that arise from $\epsilon^\prime$ do not apply to parity
conserving operators that do not contribute to the decay $K^0 \ra
\pi \pi$. In the operator basis that we are using, all the operators have
parity conserving and violating components. However, it is possible
to construct parity conserving combinations of operators just as it
is possible to construct combinations of operators with definite
isospin.

All of  the $|\Delta S|=1$ four-quark
operators that we consider
contribute to $\epsilon$ when combined with a second
$|\Delta S|=1$ vertex from the usual weak Hamiltonian.
In terms of the $K^0-\kob$ mixing matrix, each operator gives a
contribution to $\epsilon$ of the form:
\begin{equation}
|\epsilon|_i \approx {1 \over \sqrt{2}}{|{\rm Im}M_{12}|_i \over \Delta m_k}
\end{equation}
We estimate the long distance contributions to Im$M_{12}$
due to intermediate pion and eta poles \cite{dohowe}.
Using the matrix elements of Table~\ref{t: matelep} we find that there
is a cancellation between the contributions of the pion
and octet-eta poles at leading order in $SU(3)$ breaking. This
situation is unfortunate because it makes the estimates very unreliable.
For our purposes we will use the model of Ref.~\cite{dhlin}
to deal with this problem.

The contribution of each operator to $\epsilon$ is given by:
\begin{eqnarray}
|\epsilon|_{{\cal O}_i} & = & \sqrt{2} g^2 {g_8 \over M_W^2}\biggl(
{m_K \over \Delta m_K}\biggr) |{\rm Im} \lambda_i|
\biggl({M_W \over \Lambda}\biggr)^2{m_K^2 \over m_K^2-m_\pi^2}
|\xi_i|\nonumber \\
&\approx& 9.3 \ |{\rm Im} \lambda_i|
\biggl({M_W \over \Lambda}\biggr)^2|\xi_i|
\end{eqnarray}
where $g_8$ is defined in Eq.~\ref{smkpipole}, and
$\xi_i$ is given according to the model of Ref.~\cite{dhlin} by:
\begin{eqnarray}
 \xi_i &=& {1\over \sqrt{2}f_\pi^2 m_K^2}
{f_K \over f_\pi} \biggl\{<\pi^0|{\cal O}_i|\kob>
+ {<\eta_8|{\cal O}_i|\kob> \over \sqrt{3} }
\nonumber \\
&&\cdot \biggl\{\biggl({m_K^2 - m_\pi^2 \over m_K^2 - m_\eta^2}\biggr)
\biggl[(1+\xi)\cos\theta+2\sqrt{2}\rho\sin\theta\biggr]
\biggl[{f_{\eta_8}\over f_\pi}\cos\theta-\sqrt{2}
{f_{\eta_0} \over f_\pi}\sin\theta\biggr] \nonumber \\
&&+ \biggl({m_K^2 - m_\pi^2 \over m_K^2 - m_\eta^{\prime 2}}\biggr)
\biggl[(1+\xi)\sin\theta-2\sqrt{2}\rho\cos\theta\biggr]
\biggl[{f_{\eta_8}\over f_\pi}\sin\theta+\sqrt{2}
{f_{\eta_0} \over f_\pi}\cos\theta\biggr] \biggr\}\biggr\}
\label{newep}
\end{eqnarray}
We choose the parameters that Ref.~\cite{dhlin} considers more
physical: $\theta = -20^\circ$, $\xi = 0.17$, $f_{\eta 8} =
1.25 f_\pi$, $f_{\eta 0}=1.04 f_\pi$.

Once more we present separate results for the $\Delta I = 1/2, 3/2$
components of each operator in Table~\ref{t: xi}.
We emphasize again that we present
our results in this form because it is possible to construct
combinations of operators that
have definite isospin transformation properties. For the
$\Delta I = 1/2$ component, there is sensitivity to the parameters
in the model of Ref.~\cite{dhlin}. We illustrate this by presenting
results for $\rho=0.8$, $\rho = 1.2$, and for just the pion pole.
For the $\Delta I= 3/2$ component there is only a pion pole.

\begin{table}[tbh]
\centering
\caption[]{Factors $\xi_i$ for Eq.~\ref{newep}.}
\begin{tabular}{|c|c|c|c|c|} \hline
Operator & $\xi_{i,1/2} (\rho =0.8)$ & $\xi_{i,1/2} (\rho =1.2)$
& $\xi_{i,1/2}$($\pi$-only) & $\xi_{i,3/2}$ \\ \hline
% & & & & \\
${\cal O}_{qq}^{(1,1)}$  & $ -0.24 $ & $0.14$ & $-0.04$ & $0$ \\
${\cal O}_{qq}^{(8,1)}$ & $-0.41  $ & $-0.12$ & $-0.22$ & $0$ \\
${\cal O}_{qq}^{(1,3)}$ & $-0.33 $ & $-0.17$ & $-0.21$ & $0$ \\
${\cal O}_{qq}^{(8,3)}$ & $-0.16 $ & $0.71$ & $0.22$ & $0$ \\ \hline
${\cal O}_{dd}^{(1)}$ & $-0.18 $ & $0.04$ & $-0.06$ & $-0.11$ \\
${\cal O}_{ud}^{(1)}$ & $-0.06 $ & $0.1$ & $0.01$ & $0.11$ \\
${\cal O}_{dd}^{(8)}$ & $-0.23 $ & $0.06$ & $-0.07$ & $-0.15$ \\
${\cal O}_{ud}^{(8)}$ & $-0.18 $ & $-0.18$ & $-0.15$ & $0.15$ \\ \hline
${\cal O}_{qu}^{(1)}$ & $-1.8 $ & $-1.7$ & $-1.5$ & $1.5$ \\
${\cal O}_{qu}^{(8)}$ & $-0.05 $ & $0.24$ & $0.07$ & $0.15$ \\
${\cal O}_{qd}^{(1)}$ & $-4.2 $ & $-1.2$ & $-2.2$ & $0$ \\
${\cal O}_{qsd}^{(1)}$ & $-2.4 $ & $0.57$ & $-0.75$ & $-1.5$ \\
${\cal O}_{qsd}^{(8)}$ & $-0.23 $ & $0.06$ & $-0.07$ & $-0.15$ \\ \hline
${\cal O}_{qsq}^{(1)}$ & $-1.7 $ & $-2.2$ & $-1.6$ & $1.2$ \\
${\cal O}_{qsq}^{(8)}$ & $0.46 $ & $-2.1$ & $-0.65$ & $-1.3$ \\
${\cal O}_{qqs}^{(1)}$ & $0.22 $ & $-2.6$ & $-1.0$ & $-1.2$ \\
${\cal O}_{qqs}^{(8)}$ & $-1.6 $ & $-1.6$ & $-1.3$ & $1.3$ \\
${\cal O}_{qq}^{(1s)}$ & $-1.5 $ & $-4.8$ & $-2.6$ & $0$ \\
${\cal O}_{qq}^{(8s)}$ & $-1.1$ & $-3.7$ & $-2.0$ & $0$ \\ \hline
\end{tabular}
\label{t: xi}
\end{table}

\subsection{$\Lambda \ra p \pi^-$ and $A(\Lambda^0_-)$}

The starting point of the calculation is Eq.~\ref{alambda}. We study the
effect of the new physics one operator at a time and always assume that
the $\cp$ violating amplitudes are small, so that the
experimental value of the amplitudes is approximately equal to
the $\cp$ conserving amplitude. All the $\cp$ violating phases are then small
and we can write:
\begin{eqnarray}
A(\Lambda^0_-) &\approx & 3 \times 10^{-5} \nonumber \\
&+& \sum_i \biggl( 0.13 (\phi^p_1-\phi^s_1)
+0.001 (\phi^p_1-\phi^s_3) -0.0024 (\phi^p_3-\phi^s_1)\biggr)_i
\label{anew}
\end{eqnarray}
where the sum runs over all the operators in Eq.~\ref{effh}.

We carry out the calculation in the same manner as the standard model
analysis of the previous section \cite{steger}.
That is, we compute the imaginary
part of the amplitudes by taking matrix elements of each new four-quark
operator
in vacuum saturation. Further, we will not compute perturbative QCD
corrections to the effective Hamiltonian of the new physics sector.
We will also assume that the new physics does not
significantly alter the $\cp$ conserving amplitudes, but we will
comment on this later on. As discussed in
Ref.~\cite{steger}, this vacuum saturation calculation is not reliable at all,
nevertheless, we will use it for lack of anything better.

Calculating the imaginary part of the amplitudes
taking the real part from experiment as in
the previous section, we find that each operator ${\cal O}_i$ induces
the following phases:
\begin{eqnarray}
\biggl(\phi^p_1\biggr)_i &=&
-8 {G_F \over \sqrt{2}}
\biggl({M_W \over \Lambda}\biggr)^2 {\rm Im}\lambda_i
{<p \pi^-|{\cal O}_i|\Lambda>^P_1 \over 9.98 G_F m_\pi^2 }
\nonumber \\
\biggl(\phi^s_1\biggr)_i &=&
8 {G_F \over \sqrt{2}}
\biggl({M_W \over \Lambda}\biggr)^2 {\rm Im}\lambda_i
{<p \pi^-|{\cal O}_i|\Lambda>^S_1 \over 1.47G_F m_\pi^2 }
\nonumber \\
\biggl(\phi^p_3\biggr)_i &=&
8 {G_F \over \sqrt{2}}
\biggl({M_W \over \Lambda}\biggr)^2 {\rm Im}\lambda_i
{<p \pi^-|{\cal O}_i|\Lambda>^P_3  \over 0.21 G_F m_\pi^2 }
\nonumber \\
\biggl(\phi^s_3\biggr)_i &=&
-8 {G_F \over \sqrt{2}}
\biggl({M_W \over \Lambda}\biggr)^2 {\rm Im}\lambda_i
{<p \pi^-|{\cal O}_i|\Lambda>^S_3 \over 0.03 G_F m_\pi^2}
\nonumber \\
\label{hypphase}
\end{eqnarray}
The matrix elements are estimated in vacuum saturation and listed
in Table~\ref{t: matel} in the appendix. Numerically we find:
\begin{equation}
A(\Lambda^0_-) \approx  3 \times 10^{-5}
+ \biggl({M_W \over \Lambda}\biggr)^2 \sum_i {\rm Im}\lambda_i a_i
\label{newal}
\end{equation}
where the coefficients $a_i$ are listed in
Table~\ref{t: hyp}. We present two different values:
in the first column we include only the $\Delta I = 1/2$ component of each
operator, whereas in the second column we include both isospin
components. We can see from Table~\ref{t: hyp} that the $\cp$
violating asymmetry $A(\Lambda^0_-)$ is dominated by the interference
of the $s$ and $p$ waves in the $\Delta I =1/2$ amplitude, as can be
anticipated from Eq.~\ref{alambda}.

We also see from Table~\ref{t: hyp} that $a_i$ is of order one in some cases.
Eq.~\ref{newal} then tells us that a measurement of $A(\Lambda^0_-)$ at the
$10^{-4}$ level is sensitive, in principle, to new $\cp$ violating interactions
generated at a scale $\Lambda \leq 8$~TeV and is thus potentially interesting.
\begin{table}[tbh]
\centering
\caption[]{Factors $a_i$ for $A(\Lambda^0_-)_{NEW}$ in Eq.~\ref{newal}.}
\begin{tabular}{|c|c|c|} \hline
Operator &
$(a_i)_{1/2}$ & $a_i$ \\ \hline
% & & \\
${\cal O}_{qq}^{(1,1)}$ & $0.03$ & $0.03$ \\
${\cal O}_{qq}^{(8,1)}$ & $0.15$ & $0.14$ \\
${\cal O}_{qq}^{(1,3)}$ & $0.14$ & $0.14$ \\
${\cal O}_{qq}^{(8,3)}$ & $-0.15$ & $-0.15$ \\ \hline
${\cal O}_{dd}^{(1)}$ & $-0.05$ & $-0.06$ \\
${\cal O}_{ud}^{(1)}$ & $0.01$ & $0.02$ \\
${\cal O}_{dd}^{(8)}$ & $-0.06$ & $-0.08$ \\
${\cal O}_{ud}^{(8)}$ & $-0.1$ & $-0.1$ \\              \hline
${\cal O}_{qu}^{(1)}$ & $-1.3$ & $-1.1$ \\
${\cal O}_{qu}^{(8)}$ & $-0.05$ & $-0.08$ \\
${\cal O}_{qd}^{(1)}$ & $1.5$ & $1.5$ \\
${\cal O}_{qsd}^{(1)}$ & $-0.6$ & $-0.8$ \\
${\cal O}_{qsd}^{(8)}$ & $0.05$ & $0.08$ \\  \hline
${\cal O}_{qsq}^{(1)}$ & $-1.4$ & $-1.2$ \\
${\cal O}_{qsq}^{(8)}$ & $0.6$ & $0.7$ \\
${\cal O}_{qqs}^{(1)}$ & $-0.9$ & $-1.0$ \\
${\cal O}_{qqs}^{(8)}$ & $-1.1$ & $-1.0$ \\
${\cal O}_{qq}^{(1s)}$ & $1.8$ & $1.8$ \\
${\cal O}_{qq}^{(8s)}$ & $1.4$ & $1.3$
\\ \hline
\end{tabular}
\label{t: hyp}
\end{table}

We can use the constraints from $\cp$ violation in kaon decays to
place bounds on the magnitude of $A(\Lambda^0_-)$ that each of the
four quark operators can induce. In general, the bounds coming from
direct $\cp$ violation in $\epsilon^\prime$ are stronger than those
coming from $\epsilon$. However, it is necessary to consider both
because it is possible to construct parity conserving combinations of
operators that do not contribute to $K\ra \pi \pi$ amplitudes and, thus,
evade the bounds from $\epsilon^\prime$. Similarly, $\epsilon^\prime$
places stronger constraints on $\Delta I =3/2$ operators than on $\Delta I
=1/2$ operators due to the enhancement factor of $1/\omega$ in
Eq.~\ref{epprime}. To take into account these distinctions, we list
in Table~\ref{t: result} the bounds on each of the weak phases separately.
The blank entries indicate that there is no bound because the particular
operator does not contribute to that amplitude.
\begin{table}[tbh]
\centering
\caption[]{Bounds on the phases that enter $A(\Lambda^0_-)$.}
\begin{tabular}{|c|c|c|c|c|} \hline
Operator & $\phi^p_1 \times 10^{5}$ & $\phi^s_1\times 10^{5}$
 & $\phi^p_3\times 10^{5}$ & $\phi^s_3\times 10^{5}$\\ \hline
% & & & &  \\
${\cal O}_{qq}^{(1,1)}$  & $2.9$ & $-10$ & $--$ & $--$ \\
${\cal O}_{qq}^{(8,1)}$  & $9.0$ & $-10$ & $--$ & $--$ \\
${\cal O}_{qq}^{(1,3)}$  & $10$ & $-10$ & $--$ & $--$ \\
${\cal O}_{qq}^{(8,3)}$  & $-24$ & $-10$ & $--$ & $--$ \\ \hline
${\cal O}_{dd}^{(1)}$  & $5.3$ & $-10$ & $400$ & $-16$ \\
${\cal O}_{ud}^{(1)}$  & $-3.7$ & $-10$ & $400$ & $-16$ \\
${\cal O}_{dd}^{(8)}$  & $5.3$ & $-10$ & $400$ & $-16$ \\
${\cal O}_{ud}^{(8)}$  & $14$ & $-10$ & $400$ & $-16$ \\ \hline
${\cal O}_{qu}^{(1)}$  & $14$ & $-9.3$ & $400$ & $-15$ \\
${\cal O}_{qu}^{(8)}$  & $-24$ & $-10$ & $400$ & $-16$ \\
${\cal O}_{qd}^{(1)}$  & $9$ & $22$ & $--$ & $--$ \\
${\cal O}_{qsd}^{(1)}$  & $5.4$ & $2.8$ & $-400$ & $-15$ \\
${\cal O}_{qsd}^{(8)}$  & $5.3$ & $-10$ & $400$ & $-16$ \\ \hline
${\cal O}_{qsq}^{(1)}$  & $16$ & $-14$ & $400$ & $-15$ \\
${\cal O}_{qsq}^{(8)}$  & $24$ & $-2.8$ & $-400$ & $15$ \\
${\cal O}_{qqs}^{(1)}$  & $-74$ & $4.2$ & $400$ & $-15$ \\
${\cal O}_{qqs}^{(8)}$  & $14$ & $-9.3$ & $400$ & $-15$ \\
${\cal O}_{qq}^{(1s)}$  & $29$ & $22$ & $--$ & $--$ \\
${\cal O}_{qq}^{(8s)}$  & $29$ & $22$ & $--$ & $--$ \\ \hline
\end{tabular}
\label{t: result}
\end{table}

The bounds on the $p$-wave phases arise from the contributions of the
operator to $\epsilon$, and are weaker than the bounds on the $s$-wave
phases that arise from the contributions to $\epsilon^\prime$.
For the operator basis that we have been using, the bounds on the
different components are not independent. This, however, is not an
important point because there is nothing special about this operator
basis. We prefer to illustrate separately the bounds on each parity
and isospin amplitude because it
is possible to construct operators with definite parity and isospin.

\section{Two-quark Operators of Dimension six}

In addition to the four-quark operators of dimension six considered in the
previous section, there are also two-quark operators of dimension six that
can contribute to the processes under consideration \cite{buchmuller}.
These operators are the $SU(3)\times SU(2) \times U(1)$ invariant versions
of ``penguin'' operators that naively appear to be dimension five
\cite{rujula}. There are two types of operators that contribute to
$\cp$ violation in $|\Delta S|=1$ processes. The first one, in the
notation of Ref.~\cite{buchmuller} is:
\begin{equation}
{\cal O}_{dG} = (\overline{q}\sigma_{\mu\nu}\lambda^a d)\phi G^a_{\mu\nu}
\end{equation}
The operator of interest to us is obtained when the scalar
doublet, $\phi$,  takes
its vacuum expectation value. This leads to the effective Lagrangian
(with the same overall normalization that we used before and
$v \approx 246$~GeV):
\begin{eqnarray}
{\cal L}_p&=&{g^2 \over \Lambda^2}{v \over \sqrt{2}}\biggl[
\lambda_{ds}\dd\sigma_{\mu\nu}\lambda^a\biggl({1+\gamma_5 \over 2}\biggr)
sG^a_{\mu\nu} + \lambda^\star_{sd}\dd\sigma_{\mu\nu}\lambda^a\biggl(
{1-\gamma_5\over 2}\biggr)sG^a_{\mu\nu}\biggr]+{\rm h.~c.}
\nonumber \\
&\equiv & {g^2 \over \Lambda^2}{v \over \sqrt{2}}
\dd\sigma_{\mu\nu}t^a(f_{pc}+\gamma_5 f_{pv})sG^a_{\mu\nu} +{\rm h.~c.}
\label{penguinop}
\end{eqnarray}
There are also analogous operators where the Gluon field strength tensor
is replaced by field strength tensors for electroweak gauge bosons. The
matrix elements of these operators are suppressed by a power of
$\alpha = 1/137$ with respect to the gluon operator and we will,
therefore, neglect them.

\subsection{Constraint on the parity conserving coupling}

The parity conserving coupling $f_{pc}$ is constrained by the contribution
of Eq.~\ref{penguinop} to the parameter $\epsilon$. Unlike the four-quark
operators of the previous section, we cannot use vacuum saturation to compute
the matrix elements of this operator. However, this is the same
operator that arises in the Weinberg model of $\cp$ violation, and the
analysis has been carried out by Donoghue and Holstein \cite{dohowe}
using MIT bag model matrix elements. We can simply take over their
results to find:
\begin{equation}
|\epsilon|_p \approx 1.5 \times 10^{5} |\xi| \biggl({M_W\over
\Lambda}\biggr)^2|{\rm Im}f_{pc}|
\label{epconp}
\end{equation}
This contribution to $\epsilon$ is due to long distance effects as those
discussed in the previous section. In complete analogy we have introduced
the parameter $\xi$ which takes the values $\xi = 0.12$ for $\rho =0.8$
and $\xi = -0.48$ for $\rho =1.2$. We find that the sensitivity of the
result to the $SU(3)$ breaking parameters of the pole model is larger
in this case than it was for the four-quark operators.

\subsection{Constraint on the parity violating coupling}

The constraint on the parity violating coupling $f_{pv}$ comes from an
analysis of $\epsilon^\prime$. Just as we did for $f_{pc}$, we simply
take over the results of Ref.~\cite{dohowe} with a suitable identification
of the coupling. We find:
\begin{equation}
\biggl|{\epsilon^\prime \over \epsilon}\biggr|_p \approx
2.2 \times 10^5 \biggl({M_W \over \Lambda}\biggr)^2
|{\rm Im}f_{pv}|
\label{eppconp}
\end{equation}

\subsection{Contribution to $A(\Lambda^0_-)$}

Once again we use the fact that up to coupling constants, this operator
is the same one appearing in the Weinberg model of $\cp$ violation. Its
matrix elements using the MIT bag model can thus be taken from
Ref.~\cite{donpa}. We find:
\begin{eqnarray}
\phi^s_1 &\approx & 7 \times 10^4 \biggl({M_W \over \Lambda} \biggr)^2
{\rm Im}f_{pv} \nonumber \\
\phi^p_1 & \approx & -8 \times 10^4 \biggl({M_W \over \Lambda} \biggr)^2
{\rm Im}f_{pc}
\end{eqnarray}
{}From these it follows that:
\begin{equation}
A(\Lambda^0_-)_p\approx -10^4 \biggl({M_W \over \Lambda} \biggr)^2
\biggl({\rm Im}f_{pc}+0.9{\rm Im}f_{pv}\biggr)
\end{equation}
In the Weinberg model this operator appears with $f_{pv}=-f_{pc}$ and
there is a large cancellation between the two phases leading to a
smaller value for $A(\Lambda^0_-)$ than would have been obtained from
each phase individually \cite{donpa}.
In our general operator analysis, the bounds
from Eq.~\ref{epconp} and Eq.~\ref{eppconp} can be combined to obtain
(with $\xi =-0.5$):
\begin{eqnarray}
\biggl({M_W \over \Lambda} \biggr)^2 {\rm Im}f_{pc} &<& 2.7 \times 10^{-8}
\nonumber \\
\biggl({M_W \over \Lambda} \biggr)^2 {\rm Im}f_{pv} &<& 6.8 \times 10^{-9}
\end{eqnarray}
or in terms of the hyperon decay observable:
\begin{equation}
A(\Lambda^0_-)  \leq  \cases{ 3 \times 10^{-4} & Parity conserving operator \cr
6 \times 10^{-5} & Parity violating operator \cr}
\label{respen}
\end{equation}

Before ending this section we should comment on one class of two-quark
operators that we have not discussed. In the notation of Ref.~\cite{buchmuller}
it is:
\begin{equation}
{\cal O}_{qG} = i(\overline{q}\lambda^a \gm D_\nu d)\phi G^a_{\mu\nu}
\label{other}
\end{equation}
and related operators with field strength tensors for electroweak gauge
bosons instead of the gluon. These latter ones will have matrix elements
suppressed by $\alpha$ compared to Eq.~\ref{other}. We have not found a
simple way to estimate the matrix elements of this operators and for this
reason we do not discuss them in detail. We do not expect the behavior of this
type of operator to be significantly different from the others that we have
discussed.

\section{Summary and Conclusions}

The minimal standard model of electroweak interactions
is in extraordinary agreement with all experiments conducted so far,
and there is no evidence for any new particles below $100$~GeV or so.
In view of this, it is reasonable to assume that any new physics beyond
the minimal standard model is associated with a scale $\Lambda \geq M_W$,
and it is, therefore, possible to represent the low energy effects of any
such new physics with an effective Lagrangian that respects the symmetries
of the standard model.

In this paper we have studied all the $|\Delta S| =1$, $\cp$ violating,
operators that occur at dimension six. We have investigated the constraints
that exist on the couplings of these operators from the measurements of
$\epsilon$ and $\epsilon^\prime$, and estimated what their largest
contribution to $\cp$ violation in $A(\Lambda^0_-)$ could be.

The operators that we have discussed also contribute
to CP conserving and flavor changing amplitudes. We might thus worry,
that the constraints on the real part of the couplings are such, that
it is not natural for the imaginary ($\cp$ violating) part of the
couplings to attain the upper bounds allowed by the values of $\epsilon$
and $\epsilon^\prime$. We briefly address this issue in this section.

Consider the contributions to $K^0-\kob$ mixing. If we fix
Im~$\lambda_i$ to its maximum allowed value, we find that the
constraint $2{\rm Re}M_{12,i} \leq \Delta m_K$ is also satisfied if:
\begin{equation}
{\rm Re}\lambda_i \leq {{\rm Im}\lambda_i \over 2 \sqrt{2}\epsilon}
\end{equation}
Therefore, the $\cp$ conserving constraint is also satisfied if
both real and imaginary parts of the couplings are of the same size
or if the imaginary part is smaller than the real part by a factor
of $\epsilon$.
The strongest constraints on flavor changing operators
in the $\cp$ conserving case are known to come from $K^0-\kob$ mixing.
If we set the couplings to be of order one, we obtain
a lower bound on the scale of new physics $\Lambda$ requiring
that $2{\rm Re}M_{12,i} \leq \Delta m_K$. It is easy to check that
with couplings and scales satisfying this bound, the new operators do
not make any significant contributions to the real part of the amplitudes
in $K \ra \pi \pi$ or $\Lambda \ra p \pi$.
Therefore, we conclude that fixing the imaginary part of the couplings to
their maximum allowed value is not in conflict with $\cp$
conserving constraints.

In the minimal standard model we have estimated previously \cite{steger}
that $A(\Lambda^0_-)$ is of the order of a few times $10^{-5}$. For the
new physics considered in this paper we find that most of the operators
would naturally induce contributions to $A(\Lambda^0_-)$ at the $10^{-5}$
level, making them indistinguishable from the minimal standard model
(as long as precise calculations of the matrix elements are not available),
and inaccessible to the search to be conducted by E871. However, we have
also found that for certain operators, ${\cal O}_{qqs}^{(1)}$ and
${\cal O}_{dG}$, $A(\Lambda^0_-)$ could be as large as a few times
$10^{-4}$.

Given our crude estimate of the hadronic matrix elements involved,
all our numerical results should be viewed with caution. Nevertheless,
our results suggest that the search for $\cp$ violation in $A(\Lambda^0_-)$
at the $10^{-4}$ level of sensitivity that is expected for E871 is
potentially very interesting. Our results also suggest that this
measurement is complementary to the measurement of $\epsilon^\prime/
\epsilon$, in that it probes potential sources of $\cp$ violation
at a level that has not been probed by the kaon experiments. This
is particularly true for parity conserving interactions that do not
contribute to $\epsilon^\prime$ and are only constrained by $\epsilon$.

We conclude that it is possible for E871 to observe a $\cp$ violating
signal at the $10^{-4}$ level. Our study indicates that if such a signal
is observed, it would probably be evidence for physics beyond the minimal
standard model. However, a reliable determination of hadronic matrix
elements is necessary to reach any definite conclusion.

\section*{Acknowledgements}

The work of G.V. was supported in part by a DOE OJI award under contract
number DE-FG02-92ER40730. The work of X.G.He was supported in part by
DOE contract number DE-FG06-85ER40224.
G.V. thanks the Institute of Theoretical Science
at the University of Oregon for their hospitality while part of this work
was performed. We are grateful to  J. F. Donoghue and N. Deshpande for
helpful discussions.

\clearpage

\appendix

\section{Operator Analysis}

\subsection{Dimension six $|\Delta S|=1$ four-quark operators}

In Table~\ref{t: operators} we list all the four quark operators of
dimension six that change strangeness by one unit. We use the
explicitly $SU(3)\times SU(2)\times U(1)$ gauge invariant notation
of Ref.~\cite{buchmuller}. For each class of gauge invariant operator
we give the components needed for this paper.

\begin{table}[tbh]
\centering
\caption[]{Dimension six $|\Delta S|=1$ four-quark operators. We list in the
second column the gauge invariant version of the operator in the notation
of Ref.~\cite{buchmuller}, and in the third
column the $|\Delta S|=1$ components (in some cases there is more than one).}
\begin{tabular}{|c|c|c|} \hline
Operator & Ref.~\cite{buchmuller} & $|\Delta S| =1$ \\ \hline
% & & \\
${\cal O}_{qq}^{(1,1)}$ & ${\cal O}^{(1,1)}_{}$ &
${1\over 2}\dd_L \gm s_L (\uu_L\gm u_L + \dd_L \gm d_L)$ \\
${\cal O}_{qq}^{(8,1)}$ & ${\cal O}^{(8,1)}_{}$ &
${1\over 2}\dd_L \ll \gm s_L (\uu_L\ll \gm u_L + \dd_L \ll\gm d_L)$ \\
${\cal O}_{qq}^{(1,3)}$ & ${\cal O}^{(1,3)}_{}$ &
${1\over 2}(2\uu_L \gm s_L \dd_L\gm u_L - \uu_L\gm u_L\dd_L\gm s_L +
\dd_L\gm d_L\dd_L\gm s_L  )$ \\
${\cal O}_{qq}^{(8,3)}$ & ${\cal O}^{(8,3)}_{}$ &
${1\over 2}(2\uu_L \ll\gm s_L \dd_L \ll\gm u_L - \uu_L \ll\gm
u_L\dd_L\ll\gm s_L+\dd_L\ll\gm d_L\dd_L\ll\gm s_L )$ \\ \hline
${\cal O}_{dd}^{(1)}$ & ${\cal O}^{(1)}_{dd}$ &
${1\over 2}\dd_R\gm s_R \dd_R \gm d_R$ \\
${\cal O}_{ud}^{(1)}$ & ${\cal O}^{(1)}_{ud}$ &
${1\over 2}\uu_R\gm u_R \dd_R \gm s_R$ \\
${\cal O}_{dd}^{(8)}$ & ${\cal O}^{(8)}_{dd}$ &
${1\over 2}\dd_R \ll\gm s_R \dd_R \ll \gm d_R$ \\
${\cal O}_{ud}^{(8)}$ & ${\cal O}^{(8)}_{ud}$ &
${1\over 2}\uu_R \ll\gm u_R \dd_R \ll\gm s_R$ \\ \hline
${\cal O}_{qu}^{(1)}$ & ${\cal O}^{(1)}_{qu}$ &
$\dd_L u_R \uu_R s_L$ \\
${\cal O}_{qu}^{(8)}$ & ${\cal O}^{(8)}_{qu}$ &
$\dd_L\ll u_R \uu_R \ll s_L$ \\
${\cal O}_{qd}^{(1)}$ & ${\cal O}^{(1)}_{qd}$ &
$\uu_L s_R\dd_R u_L +\dd_L s_R\dd_R d_L$ \\
${\cal O}_{qsd}^{(1)}$ & ${\cal O}^{(1)}_{qd}$ &
$\dd_L d_R \dd_R s_L$ \\
${\cal O}_{qsd}^{(8)}$ & ${\cal O}^{(8)}_{qd}$ &
$\dd_L\ll d_R \dd_R \ll s_L$ \\ \hline
${\cal O}_{qsq}^{(1)}$ & ${\cal O}^{(1)}_{qq}$ &
$-\uu_Rs_L\dd_Ru_L $ \\
${\cal O}_{qsq}^{(8)}$ & ${\cal O}^{(8)}_{qq}$ &
$-\uu_R \ll s_L\dd_R \ll u_L $ \\
${\cal O}_{qqs}^{(1)}$ & ${\cal O}^{(1)}_{qq}$&
$\dd_R s_L \uu_R u_L$ \\
${\cal O}_{qqs}^{(8)}$ & ${\cal O}^{(8)}_{qq}$&
$\dd_R\ll s_L \uu_R \ll u_L$ \\
${\cal O}_{qq}^{(1s)}$ & ${\cal O}^{(1)}_{qq}$&
$\uu_L u_R \dd_L s_R - \dd_L u_R \uu_L s_R $ \\
${\cal O}_{qq}^{(8s)}$ & ${\cal O}^{(8)}_{qq}$&
$\uu_L \ll u_R \dd_L\ll s_R - \dd_L\ll u_R \uu_L\ll s_R $
\\ \hline
\end{tabular}
\label{t: operators}
\end{table}

\clearpage
\subsection{Isospin decomposition}

For convenience we provide the isospin decomposition of the
four-quark operators in Table~\ref{t: isospin}.

\begin{table}[tbh]
\centering
\caption[]{Isospin decomposition of four-quark operators.}
\begin{tabular}{|c|c|c|} \hline
Operator &  $\Delta I = 1/2$  & $\Delta I = 3/2$ \\ \hline
% & & \\
$3\uu s \dd u$ &
$2 \uu s \dd u - \dd s \uu u + \dd s \dd d $ &
$ \uu s \dd u + \dd s \uu u - \dd s \dd d $ \\
$3\dd s \uu u $ &
$2 \dd s \uu u - \uu s \dd u + \dd s \dd d $ &
$ \uu s \dd u + \dd s \uu u - \dd s \dd d $ \\
$3\dd s \dd d $ &
$ \uu s \dd u + \dd s \uu u + 2\dd s \dd d $ &
$- \uu s \dd u - \dd s \uu u + \dd s \dd d $ \\ \hline
\end{tabular}
\label{t: isospin}
\end{table}

\subsection{Matrix elements for $\Lambda \ra p \pi^-$ in vacuum saturation}

We use the normalization in which $f_\pi = 93$~MeV,
and neglect $m_{u,d}/m_s$. In terms of
\begin{equation}
B_0 \equiv  {m_\pi^2 \over m_u + m_d}={m_K^2 \over m_s +m_u}
\approx 11\  m_\pi
\end{equation}
we find:
\begin{eqnarray}
<p \pi^- |\dd \gamma_\mu \gamma_5 u \uu \gamma^\mu s|\Lambda> &\equiv & M_V =
i \sqrt{2} f_\pi (M_\Lambda - M_P)\sqrt{3 \over 2}\ff_p\Psi_\Lambda
\nonumber \\
<p \pi^- |\dd \gamma_\mu \gamma_5 u \uu \gamma^\mu \gamma_5s|\Lambda>
&\equiv & M_A =
-i {2 f_\pi f_K m^2_\pi \over m_K^2 - m^2_\pi}(-13.3)\ff_p \gamma_5\Psi_\Lambda
\nonumber \\
<p \pi^- |\dd \gamma_5 u \uu  s|\Lambda> &=& {B_0^2 \over m_K^2} M_V
\nonumber \\
<p \pi^- |\dd \gamma_5 u \uu  \gamma_5 s|\Lambda> &=&- {B_0^2 \over m_K^2} M_A
\end{eqnarray}
We list the matrix element for each operator using vacuum saturation in
Table~\ref{t: matel}

\clearpage
\begin{table}[tbh]
\centering
\caption[]{Matrix elements in $\Lambda \ra p \pi^-$.}
\begin{tabular}{|c|c|c|} \hline
Operator &  $\Delta I = 1/2$  & $\Delta I = 3/2$ \\ \hline
% & & \\
${\cal O}_{qq}^{(1,1)}$ &
${1 \over 8N}(M_A-M_V)$ & 0 \\
${\cal O}_{qq}^{(8,1)}$ &
${1\over 4}\biggl(1-{1\over N^2}\biggr)(M_A-M_V)$ & 0 \\
${\cal O}_{qq}^{(1,3)}$ &
${1\over 4}\biggl(1-{1\over 2N}\biggr)(M_A-M_V) $ & 0 \\
${\cal O}_{qq}^{(8,3)}$ &
${1\over 4}\biggl({1\over N^2}-1\biggr)(M_A-M_V)$ & 0 \\ \hline
${\cal O}_{dd}^{(1)}$ &
${1\over 24}\biggl(1+{1\over N}\biggr)(M_A+M_V)$ &
$-{1\over 24}\biggl(1+{1\over N}\biggr)(M_A+M_V)$ \\
${\cal O}_{ud}^{(1)}$ &
${1\over 24}\biggl({2\over N}-1\biggr)(M_A+M_V)$ &
${1\over 24}\biggl(1+{1\over N}\biggr)(M_A+M_V)$ \\
${\cal O}_{dd}^{(8)}$ &
${1\over 12}\biggl(1-{1\over N^2}\biggr)(M_A+M_V)$ &
$-{1\over 12}\biggl(1-{1\over N^2}\biggr)(M_A+M_V)$ \\
${\cal O}_{ud}^{(8)}$ &
${1\over 6}\biggl(1-{1\over N^2}\biggr)(M_A+M_V)$ &
${1\over 12}\biggl(1-{1\over N^2}\biggr)(M_A+M_V)$ \\ \hline
${\cal O}_{qu}^{(1)}$ &
${1\over 12}\biggl(2{B_0^2 \over m_K^2}
(M_V+M_A)+{1\over 2N}(M_V-M_A)\biggr)$ &
${1\over 12}\biggl({B_0^2 \over m_K^2}
(M_V+M_A)-{1\over 2N}(M_V-M_A)\biggr)$ \\
${\cal O}_{qu}^{(8)}$ &
${1\over 12}\biggl(1-{1\over N^2}\biggr)(M_V-M_A)$ &
${1\over 12}\biggl({1\over N^2}-1\biggr)(M_V-M_A)$  \\
${\cal O}_{qd}^{(1)}$ &
$-{B_0^2 \over 4 m_K^2}(M_V-M_A) $ & 0 \\
${\cal O}_{qsd}^{(1)}$ &
${1 \over 12}\biggl[{B_0^2 \over m_K^2}
(M_V+M_A)-{1\over 2N}(M_V-M_A)\biggr]$ &
${1 \over 12}\biggl[-{B_0^2 \over m_K^2}
(M_V+M_A)+{1\over 2N}(M_V-M_A)\biggr]$ \\
${\cal O}_{qsd}^{(8)}$ &
${1\over 12}\biggl({1\over N^2}-1\biggr)(M_V-M_A)$ &
$-{1\over 12}\biggl({1\over N^2}-1\biggr)(M_V-M_A)$ \\ \hline
${\cal O}_{qsq}^{(1)}$ &
${1\over 12}\biggl(2 + {1\over 2N}\biggr){B_0^2 \over m_K^2}(M_V+M_A)$ &
${1\over 12}\biggl(1- {1\over 2N}\biggr){B_0^2 \over m_K^2}(M_V+M_A) $ \\
${\cal O}_{qsq}^{(8)}$ &
$-{1\over 12}\biggl(1- {1\over N^2}\biggr){B_0^2 \over m_K^2}(M_V+M_A)$ &
${1\over 12}\biggl(1- {1\over N^2}\biggr){B_0^2 \over m_K^2}(M_V+M_A) $ \\
${\cal O}_{qqs}^{(1)}$ &
${1 \over 12}\biggl(1+{1\over N}\biggr){B_0^2 \over m_K^2}(M_V+M_A)$ &
$-{1 \over 12}\biggl(1-{1\over 2N}\biggr){B_0^2 \over m_K^2}(M_V+M_A)$ \\
${\cal O}_{qqs}^{(8)}$ &
${1 \over 6}\biggl(1-{1\over N^2}\biggr){B_0^2 \over m_K^2}(M_V+M_A)$ &
${1 \over 12}\biggl(1-{1\over N^2}\biggr){B_0^2 \over m_K^2}(M_V+M_A)$ \\
${\cal O}_{qq}^{(1s)}$ &
$-{1\over 4}\biggl(1+{1\over 2N}\biggr){B_0^2 \over m_K^2}(M_V-M_A)$ & 0 \\
${\cal O}_{qq}^{(8s)}$ &
$-{1\over 4}\biggl(1-{1\over N^2}\biggr){B_0^2 \over m_K^2}(M_V-M_A)$ & 0
\\ \hline
\end{tabular}
\label{t: matel}
\end{table}

\clearpage
\subsection{Matrix elements for $\kob\ra \pi^+\pi^-$
in vacuum saturation}

In vacuum saturation we find:
\begin{eqnarray}
V_1\  \equiv \ <\pi^-|\dd\gm\gamma_5 u |0><\pi^+|\uu \gm s|\kob> &= &
-i\sqrt{2}f_\pi(m_K^2-m_\pi^2) \nonumber \\
V_2 \ \equiv \ <\pi^+\pi^-|\uu \gm u|0><0|\dd \gm\gamma_5 s|\kob> &=&
-<\pi^+\pi^-|\dd \gm d|0><0|\dd \gm\gamma_5 s|\kob>
\nonumber \\ &=& 0 \nonumber \\
S_1\ \equiv \ <\pi^-|\dd\gamma_5 u |0><\pi^+|\uu  s|\kob> &= &
-i\sqrt{2}f_\pi
B_0^2 \biggl(1 +2 {m_\pi^2 \over \Lambda_\chi^2}\biggr)  \nonumber \\
S_2 \ \equiv \ <\pi^+\pi^-|\uu  u|0><0|\dd \gamma_5 s|\kob> &=&
<\pi^+\pi^-|\dd  d|0><0|\dd \gamma_5 s|\kob>
\nonumber \\ &=&
-i\sqrt{2}f_\pi
B_0^2 \biggl(1 +2 {m_K^2 \over \Lambda_\chi^2}\biggr)
\end{eqnarray}
We have to introduce momentum dependent terms in the last two expressions
because the leading terms cancel in the difference $S_1 - S_2$. The
scale of chiral symmetry breaking $\Lambda_\chi \approx 1$~GeV can be
related to the ratio $f_K/f_\pi$ \cite{donbook}. We list the matrix
elements for each operator in Table~\ref{t: matelkpp}.
\clearpage
\begin{table}[tbh]
\centering
\caption[]{Matrix elements in $\kob\ra\pi^+\pi^-$.}
\begin{tabular}{|c|c|c|} \hline
Operator &  $A_0$  & $A_2/\sqrt{2}$ \\ \hline
% & & \\
${\cal O}_{qq}^{(1,1)}$ &
${1\over 8N}(V_2-V_1)$ & 0 \\
${\cal O}_{qq}^{(8,1)}$ &
${1\over 4}\biggl(1-{1\over N^2}\biggr)(V_2-V_1)$ & 0 \\
${\cal O}_{qq}^{(1,3)}$ &
${1\over 8}\biggl(2-{1\over N}\biggr)(V_2-V_1) $ & 0 \\
${\cal O}_{qq}^{(8,3)}$ &
$-{1\over 4}\biggl(1-{1\over N^2}\biggr)(V_2-V_1)$ & 0 \\ \hline
${\cal O}_{dd}^{(1)}$ &
${1\over 24}\biggl(1+{1\over N}\biggr)(V_1-V_2)$ &
$-{1\over 24}\biggl(1+{1\over N}\biggr)(V_1+2V_2)$ \\
${\cal O}_{ud}^{(1)}$ &
${1\over 24}\biggl({2\over N}-1\biggr)(V_1-V_2)$ &
${1\over 24}\biggl(1+{1\over N}\biggr)(V_1+2V_2)$ \\
${\cal O}_{dd}^{(8)}$ &
${1\over 12}\biggl(1-{1\over N^2}\biggr)(V_1-V_2)$ &
$-{1\over 12}\biggl(1-{1\over N^2}\biggr)(V_1+2V_2)$ \\
${\cal O}_{ud}^{(8)}$ &
${1\over 6}\biggl(1-{1\over N^2}\biggr)(V_1-V_2)$ &
${1\over 12}\biggl(1-{1\over N^2}\biggr)(V_1+2V_2)$ \\ \hline
${\cal O}_{qu}^{(1)}$ &
${1\over 12}\biggl(2S_1+{1\over 2N}(V_1+V_2)\biggr)$ &
${1\over 12}\biggl(S_1-{1\over 2N}(V_1-2V_2)\biggr)$ \\
${\cal O}_{qu}^{(8)}$ &
${1\over 12}\biggl(1-{1\over N^2}\biggr)(V_1+V_2)$ &
$-{1\over 12}\biggl(1-{1\over N^2}\biggr)(V_1-2V_2)$  \\
${\cal O}_{qd}^{(1)}$ &
${1 \over 4}(S_2-S_1) $ & 0 \\
${\cal O}_{qsd}^{(1)}$ &
${1 \over 12}\biggl((S_1-3S_2)-{1\over 2N}(V_1+V_2)\biggr)$ &
${1 \over 12}\biggl(-S_1+{1\over 2N}(V_1-2V_2)\biggr)$ \\
${\cal O}_{qsd}^{(8)}$ &
$-{1\over 12}\biggl(1-{1\over N^2}\biggr)(V_1+V_2)$ &
${1\over 12}\biggl(1-{1\over N^2}\biggr)(V_1-2V_2)$ \\ \hline
${\cal O}_{qsq}^{(1)}$ &
${1\over 12}\biggl(2S_1 + {1\over 2N}(S_1-3S_2)\biggr)$ &
${1\over 12}\biggl(1- {1\over 2N}\biggr)S_1 $ \\
${\cal O}_{qsq}^{(8)}$ &
${1\over 12}\biggl(1- {1\over N^2}\biggr)(S_1-3S_2)$ &
$-{1\over 12}\biggl(1- {1\over N^2}\biggr)S_1 $ \\
${\cal O}_{qqs}^{(1)}$ &
${1 \over 12}\biggl(\biggl(1+{1\over N}\biggr)S_1-3S_2\biggr)$ &
$-{1\over 12}\biggl(1- {1\over 2N}\biggr)S_1 $ \\
${\cal O}_{qqs}^{(8)}$ &
${1 \over 6}\biggl(1-{1\over N^2}\biggr)S_1$ &
${1 \over 12}\biggl(1-{1\over N^2}\biggr)S_1$ \\
${\cal O}_{qq}^{(1s)}$ &
$-{1\over 4}\biggl(1+{1\over 2N}\biggr)(S_1-S_2)$ & 0 \\
${\cal O}_{qq}^{(8s)}$ &
$-{1\over 4}\biggl(1-{1\over N^2}\biggr)(S_1-S_2)$ & 0
\\ \hline
\end{tabular}
\label{t: matelkpp}
\end{table}

\clearpage
\subsection{Matrix elements for $\kob \ra\pi^0,\ \eta_8,\ \eta_0$
in vacuum saturation}

The matrix elements for the $\kob \ra \pi^0,\ \eta_8,\ \eta_0$
transition in the standard model are, to lowest order in chiral perturbation
theory:
\begin{eqnarray}
-i<\pi^0|H_W|\kob(q)> &=& -i2 {g_8 \over f_\pi^2}{q^2\over \sqrt{2}}\nonumber
\\
-i<\eta_8|H_W|\kob(q)> &=& -i2 {g_8 \over f_\pi^2}{q^2\over \sqrt{6}}
\nonumber \\
-i<\eta_0|H_W|\kob(q)> &=&i 2{g_8 \over f_\pi^2}{2q^2\over \sqrt{3}}
\label{smkpipole}
\end{eqnarray}
where $g_8 \approx 7.8\times 10^{-8}f_\pi^2 \approx 10^{-13} M_W^2$.

For the matrix elements of the four quark operators we use vacuum
saturation and $U(3)$ symmetry to include the $\eta$-singlet. The results
are listed in Table~\ref{t: matelep} where we have factored out a
common $\sqrt{2}f_\pi^2 m_K^2$.
For all the operators in Table~\ref{t: matelep} we have
$<\eta_0|{\cal O}|\kob> = \sqrt{2}<\eta_8|{\cal O}|\kob> $.

\clearpage

\begin{table}[tbh]
\centering
\caption[]{Matrix elements in $\kob\ra\pi^0,\ \eta_8,\ \eta_0$ An overall
$\sqrt{2}f_\pi^2m_K^2$ has been factored from all the entries in the table.}
\begin{tabular}{|c|c|c|c|} \hline
Operator &  $<\pi^0|{\cal O}_1|\kob> $ & $\sqrt{3}<\eta_8|{\cal O}_1|\kob>$ &
$<\pi^0|{\cal O}_3|\kob> $
 \\ \hline
% & & &   \\
${\cal O}_{qq}^{(1,1)}$ & $-{1\over 8N} $ &
${1\over 8 }\biggl(2+{1\over N}\biggr) $ & $0$ \\
${\cal O}_{qq}^{(8,1)}$ &
$-{1\over 4}\biggl(1-{1\over N^2}\biggr)$ &
${1\over 4 }\biggl(1-{1\over N^2}\biggr) $ & $0$ \\
${\cal O}_{qq}^{(1,3)}$ &
${1\over 8}\biggl({1\over N}-2\biggr) $ &
${3\over 8N} $ & $0$ \\
${\cal O}_{qq}^{(8,3)}$ &
${1\over 4}\biggl(1-{1\over N^2}\biggr)$ &
${3\over 4}\biggl(1-{1\over N^2}\biggr)$& $0$ \\ \hline
${\cal O}_{dd}^{(1)}$ &
$-{1\over 24}\biggl(1+{1\over N}\biggr)$ &
${1\over 8}\biggl(1+{1\over N}\biggr)$&
$-{1\over 12}\biggl(1+{1\over N}\biggr)$ \\
${\cal O}_{ud}^{(1)}$ &
${1\over 24}\biggl(1-{2\over N}\biggr)$ &
${1\over 8}$&
${1\over 12}\biggl(1+{1\over N}\biggr)$ \\
${\cal O}_{dd}^{(8)}$ &
$-{1\over 12}\biggl(1-{1\over N^2}\biggr)$ &
${1\over 4}\biggl(1-{1\over N^2}\biggr)$&
$-{1\over 6}\biggl(1-{1\over N^2}\biggr)$ \\
${\cal O}_{ud}^{(8)}$ &
$-{1\over 6}\biggl(1-{1\over N^2}\biggr)$ & 0&
${1\over 6}\biggl(1-{1\over N^2}\biggr)$ \\ \hline
${\cal O}_{qu}^{(1)}$ &
$-{1\over 12}\biggl(2\rrr-{1\over 2N}\biggr)$ &
${1\over 8N}$&
${1\over 12}\biggl(2\rrr+{1\over N}\biggr)$ \\
${\cal O}_{qu}^{(8)}$ &
${1\over 12}\biggl(1-{1\over N^2}\biggr)$ &
${1\over 4}\biggl(1-{1\over N^2}\biggr)$ &
${1\over 6}\biggl(1-{1\over N^2}\biggr)$ \\
${\cal O}_{qd}^{(1)}$ &
$-{1 \over 4}\rrr $ &
${1\over 4}\biggl({1\over N}+\rrr\biggr)$& $0$ \\
${\cal O}_{qsd}^{(1)}$ &
${1 \over 12}\biggl(-\rrr-{1\over 2N}\biggr) $ &
$-{1 \over 4}\biggl(-\rrr-{1\over 2N}\biggr)$&
${1 \over 6}\biggl(-\rrr-{1\over 2N}\biggr) $ \\
${\cal O}_{qsd}^{(8)}$ &
$-{1\over 12}\biggl(1-{1\over N^2}\biggr)$ &
${1\over 4}\biggl(1-{1\over N^2}\biggr)$&
$-{1\over 6}\biggl(1-{1\over N^2}\biggr)$ \\ \hline
${\cal O}_{qsq}^{(1)}$ &
$-{1\over 12}\biggl(2+{1\over 2 N}\biggr)\rrr $ &
$-{1\over 8N}\rrr $&
${1\over 6}\biggl(1-{1\over 2 N}\biggr)\rrr$ \\
${\cal O}_{qsq}^{(8)}$ &
$-{1\over 12}\biggl(1- {1\over N^2}\biggr)\rrr$ &
$-{1\over 4}\biggl(1- {1\over N^2}\biggr)\rrr $&
$-{1\over 6}\biggl(1- {1\over N^2}\biggr)\rrr$ \\
${\cal O}_{qqs}^{(1)}$ &
$-{1 \over 12}\biggl(1+{1\over N}\biggr)\rrr  $ &
$-{1 \over 4}\rrr  $&
$-{1 \over 6}\biggl(1-{1\over 2N}\biggr)\rrr $ \\
${\cal O}_{qqs}^{(8)}$ &
$-{1 \over 6}\biggl(1-{1\over N^2}\biggr)\rrr $ & 0&
${1 \over 6}\biggl(1-{1\over N^2}\biggr)\rrr$ \\
${\cal O}_{qq}^{(1s)}$ &
$-{1\over 4}\biggl(1+{1\over 2N}\biggr)\rrr $ &
$-{1\over 4}\biggl(1+{1\over 2N}\biggr)\rrr $ & $0$ \\
${\cal O}_{qq}^{(8s)}$ &
$-{1\over 4}\biggl(1-{1\over N^2}\biggr)\rrr $ &
$-{1\over 4}\biggl(1-{1\over N^2}\biggr)\rrr $  & $0$ \\ \hline
\end{tabular}
\label{t: matelep}
\end{table}

\end{document}